\documentclass[%
 reprint,
 amsmath,amssymb,
 aps,
floatfix,
]{revtex4-2}

\usepackage{graphicx}
\usepackage{dcolumn}
\usepackage{bm}
\usepackage{gensymb}
\usepackage{subcaption}
\usepackage{gensymb}  
\usepackage{amsmath}
\usepackage{amsfonts}
\usepackage{amssymb}
\usepackage{makeidx}
\usepackage{graphicx}
\usepackage{layout}
\usepackage{xfrac}
\usepackage{xcolor}
\usepackage{tabularx}
\usepackage[english]{babel}
\usepackage{float}
\usepackage{appendix}
\usepackage{siunitx}
\begin{document}

\preprint{APS/123-QED}

\title{Spatial Dispersion Induced Birefringence in a Silicon Beamsplitter at 2000\,nm for Future Gravitational Wave Detectors}

\author{Alex Adam, Carl Blair, Chunnong Zhao}
\affiliation{%
 OzGrav, University of Western Australia.
}%

\date{\today}

\begin{abstract}
Cryogenic silicon test masses could reduce thermal noise and thermal distortions in the next generation of gravitational wave detectors. Birefringence is an effect that can degrades the sensitivity of a detector and is of greater concern in silicon as it is a crystalline material. This is particularly problematic for the beam splitter.  We measure the birefringence in a \textlangle\num{100}\textrangle\space float zone silicon optic oriented at \ang{45} to the incident light as in the case of a beamsplitter. There is inherent birefringence due to the spatial dispersion effect. We observe that the birefringence varied between \numlist{3.44(0.12)e-7; 1.63(0.05)e-7} and estimate the birefringence along the \textlangle\num{110}\textrangle\space axis to be \num{1.64(0.05)e-6} at \qty{2000}{\nm}. We conclude that spatial dispersion induced birefringence does not preclude silicon from use as a beamsplitter substrate.
\end{abstract}

\maketitle


\section{\label{sec:level1}Introduction}

Gravitational wave detectors have proved capable of observing gravitational waves from the mergers of compact binaries\cite{Abbott2016}. The second generation of gravitational wave detectors will be replaced in the next decade by a third generation taking advantage of emerging technologies\cite{EinsteinTelescope}. The limiting noise sources of current gravitational wave detectors are: quantum shot-noise at high frequencies (\qty{>100}{\Hz}), thermal noise in the centre of the detection band (\qty{\approx 100}{\Hz}) and, a combination of quantum radiation pressure noise and seismic noise and related phenomena at low frequencies\cite{Abbott2015} ($<$10's\,Hz).  Therefore reducing quantum noise and thermal noise will improve the detector sensitivity over much of the detection band. 

Future detectors will reduce quantum shot noise by increasing the circulating optical power in the arms of the detectors and increasing the level of squeezed vacuum injection\cite{voyager, EinsteinTelescope}. However, increasing the arm power will increase the optical power absorbed by the core optics which creates thermal distortion due to thermo-elastic and thermo-optic effects. Quantum squeezing can reduce both quantum shot noise and quantum radiation pressure noise; however, this technique is susceptible to optical loss. The materials used in future gravitational wave detectors will need to have good thermal properties to reduce thermal distortions and low optical absorption.

To combat thermal distortion new mirror substrate materials and cryogenic temperatures are being explored. The gravitational wave detector KAGRA\cite{AkutsuT2021OoKD}, for instance, uses cryogenic sapphire test masses. The core optics in the advanced LIGO and advanced Virgo detectors use fused silica substrates\cite{Abbott2015, AgathosM2015TAVd} which were chosen for their low absorption at \qty{1064}{\nm} and high mechanical Q-factor. Fused silica loses its high Q-factor at cryogenic temperatures which increases thermal noise. Fused silica also has poor thermal conductivity, which increases thermal distortion. These limitations rule out silica for use in future cryogenic detectors.

Future cryogenic detectors require changing the substrate to a material with suitable properties for high-power, cryogenic operation. These properties include: a high quality factor to reduce substrate thermal noise and low optical loss to improve squeezing performance. Additionally, to avoid distortions at high optical power, the substrate must have: high thermal conductivity, low thermal expansion, and a low thermo-optic coefficient. Crystalline silicon is a promising material with many of these properties. Silicon has: a Q factor of \num{e8} at cryogenic temperatures\cite{highQ}, high thermal conductivity of \qty{137}{\W\m^{-1}\K^{-1}}, and a coefficient of thermal expansion of \qty{0}{\K^{-1}} at two temperatures (\qtylist{18;123}{\K}) \cite{thermalexpansion}. At \qtylist{18;123}{\K}, silicon has thermo-optic coefficients of \qtylist{e-6;e-4}{\K^{-1}} respectively\cite{thermooptic}. However, silicon has a high absorption at \qty{1064}{\nm}, requiring a change in wavelength.

There are two candidate laser wavelengths for a future silicon detector, \qty{1550}{\nm} and \qty{2000}{\nm}. Compared to \qty{2000}{\nm}, \qty{1550}{\nm} light has lower absorption loss in silicon as well as more mature technology\cite{Degallaix:13, siliconabsorption}. The \qty{2000}{\nm} light has lower absorption loss in the proposed high reflective coatings\cite{aSiliconCoating} as well as reduced scattering loss. 
A wavelength close to \qty{2000}{\nm} was selected by the LIGO Voyager white paper which concluded that coating absorption loss will be the most important factor\cite{voyager}.  

Although silicon has a lower absorption than fused silica it has properties that require further investigation. Of concern is the birefringence in silicon due to spatial dispersion\cite{pastrnak, CHU2002174, NewSpatialDispersion}.  Birefringence impacts gravitational wave detector sensitivity in several ways. One such impact is optical loss which reduces the sensitivity by increasing shot noise and degrading quantum squeezing\cite{Kruger}. Test mass birefringence also complicates control of the KAGRA gravitational wave detectors\cite{kagra}. Birefringence in silicon must be understood before it can be used in a gravitational wave detector.

\begin{figure}
    \centering
    \includegraphics[width=0.9\linewidth]{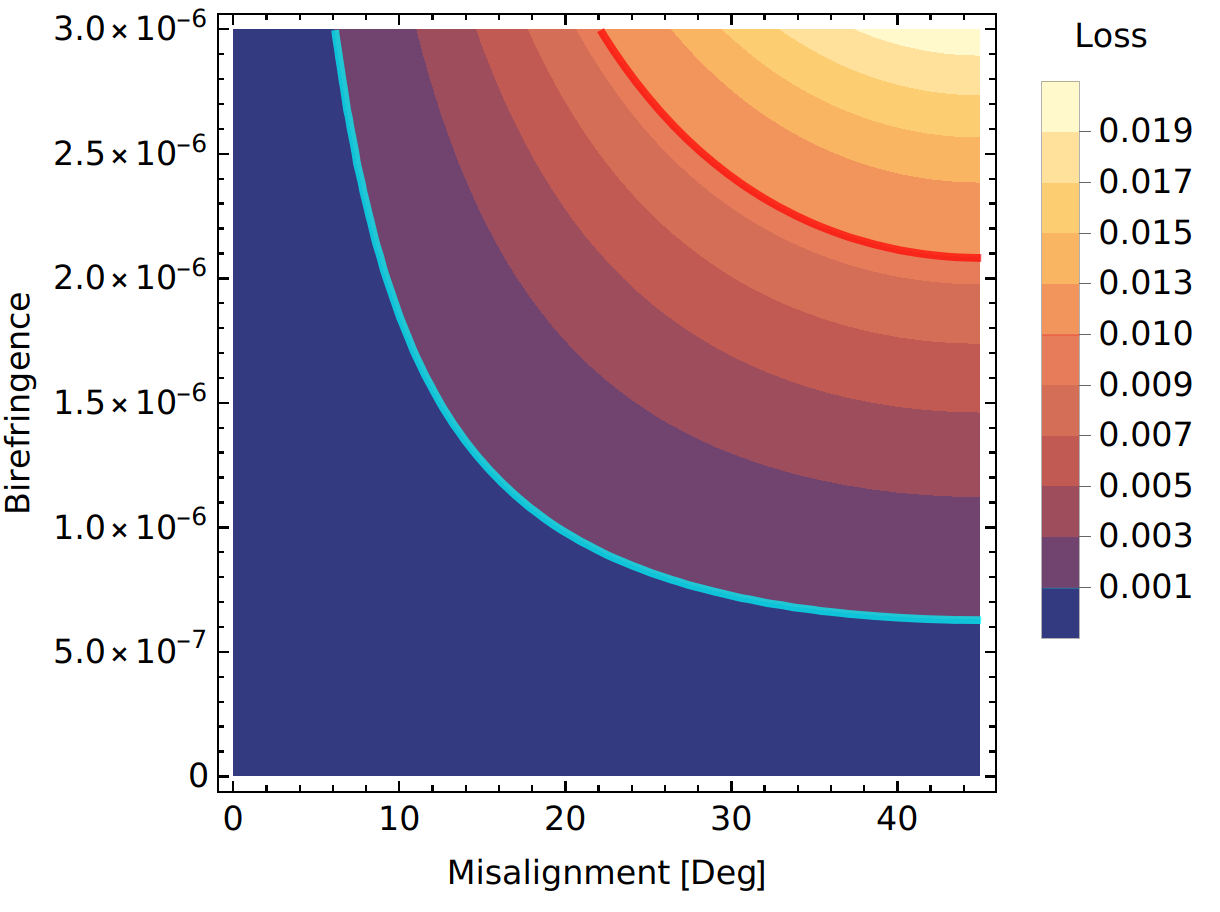}
    \caption{Optical loss from depolarisation in a silicon beamsplitter 6\,cm thick at \qty{2000}{\nm}. The cyan contour shows \qty{0.1}{\%} loss and the red shows \qty{1}{\%} loss. This does not take account of other birefringence related effects and therefore should be viewed as an indicative upper limit. The thickness of the beamsplitter is based on the parameters for the aLIGO beamsplitter \cite{Abbott2015}. }
    \label{fig:loss}
\end{figure}

It is estimated that the level of acceptable birefringence for Einstein telescope is \num{e-8}  in the case of an optic that cannot have its optic axes aligned to the input polarisation.  In the case of a well-aligned optic the limit is \num{e-7} \cite{Kruger} .  
This birefringence requirement corresponds to a 1\% optical loss limit.
If we set an optical loss requirement of \qty{0.1}{\%} from beamsplitter birefringence this produces the limit on birefringence and misalignment shown as the cyan line in Figure~\ref{fig:loss}. We assume a stress birefringence level of \num{4e-8} which is consistent with our measurements of the background birefringence below and model this as a maximally misaligned waveplate placed before the optic.

Here we investigate the suitability of using silicon as a beam-splitter where spatial dispersion can not be avoided.  We measure the birefringence in a \textlangle\num{100}\textrangle\space float zone silicon test mass with \ang{45} incident light using a \qty{2000}{\nm} laser and photoelastic modulator. We show that the birefringence in a silicon beamsplitter will be dominated by spatial dispersion, however, the level is low enough that silicon is not ruled out as a future detector beamsplitter material.

\section{Spatial Dispersion}

Birefringence is an optical anisotropy in which a material has optical axes with different refractive indices. Birefringence will generally cause linearly polarised light to become elliptically polarised depending on the alignment of the input polarisation to these optical axes. The most commonly observed form of birefringence, referred to as gyrotropy, is due to an asymmetric crystal structure. This is the origin of birefringence in crystals such as sapphire and calcite.

Although silicon is a cubic crystal and therefore non-gyrotropic there will be some background birefringence due to stress, crystal defects, impurities etc. An anisotropic term  appears in the dielectric tensor of cubic crystals when higher order effects are take into account. The crystal becomes optically heptaxial with optical axes aligned with the face normals and corner diagonals. This effect is referred to as "spatial dispersion” since the assumption of spatial dispersion allows us to derive the observed properties.

Spatial dispersion can be understood by comparing it to frequency dispersion. To describe the phase shift imparted on a time varying electric field when it passes through a material we cannot look at amplitude $E$ at a single moment. 
We look at the field over all time $t$, in other words in the frequency domain. This is equivalent to taking the Fourier transform\cite{SpatialDispersion}: 
\begin{equation}
    E(\boldsymbol{r},\omega) = \frac{1}{2\pi}\int^{\infty}_{-\infty}E(\boldsymbol{r},t)e^{i\omega t} dt,
\end{equation}
where $r$ is the spatial co-ordinate and $\omega$ the radial frequency.
Likewise, to determine the spatial dispersion effect we cannot look at the electric field at a single location, but instead take the effect of the electric field at all locations into account. 
This is the same as taking the spatial Fourier transform so the dielectric tensor now has a wave vector dependence\cite{SpatialDispersion}: 
\begin{equation}
    E(\boldsymbol{k},\omega) = \frac{1}{(2\pi)^4}\int^{\infty}_{-\infty}\int^{\infty}_{-\infty}E(\boldsymbol{r},t)e^{-i(\boldsymbol{k}\cdot\boldsymbol{r}-\omega t)} d\boldsymbol{r}dt,
\end{equation}

where $k$ is the wave vector.
Crystals have a regularly structure lattice which modulates the polarisability of the crystal at certain spacial frequencies. The effect of spatial dispersion is that cubic crystals are no longer isotropic and will exhibit a weak anisotropy. If the frequency of light is far from any electronic transitions, then this anisotropy is proportional to the inverse square of the wavelength; this provides additional support for using a longer wavelength in gravitational wave detectors. We express the spatial dispersion effect with an expansion of the dielectric tensor in terms of powers of the wave vector. Each power of the wave vector is multiplied by a tensor to account for the symmetry at each length scale\cite{SpatialDispersion}:
\begin{equation}
    \epsilon(\omega, \boldsymbol{k})_{ij} = \epsilon(\omega)_{ij} + i\gamma(\omega)_{ijl}k_l + \alpha (\omega)_{ijlm}k_lk_m + \dots
\end{equation}
or using the reciprocal of the tensor:
\begin{equation}
     \epsilon^{-1}(\omega, \boldsymbol{k})_{ij} = \epsilon^{-1}(\omega)_{ij} + i\delta(\omega)_{ijl}k_l + \beta (\omega)_{ijlm}k_lk_m + \dots
\end{equation}
The zeroth-power of the wave vector is the isotropic term, $\epsilon(\omega)$, which accounts for the averaged effect in all directions. The first power ($\gamma$ and $\delta$) accounts for gyrotropy.  In cubic crystals this term is zero. An important result occurs when we consider the quadratic term which has a tensors, $\alpha$ and $\beta$\cite{SpatialDispersion}. In cubic crystals the tensor $\beta$ introduces asymmetry and takes the form:
\begin{equation}
    \beta  =  
    \begin{pmatrix}
    \beta_{11}&\beta_{12}&\beta_{12}&0&0&0\\
    \beta_{12}&\beta_{11}&\beta_{12}&0&0&0\\
    \beta_{12}&\beta_{12}&\beta_{11}&0&0&0\\
    0&0&0&\beta_{44}&0&0\\
    0&0&0&0&\beta_{44}&0\\
    0&0&0&0&0&\beta_{44}\\
    \end{pmatrix}  
\end{equation}
The direction dependence of the normalised magnitude of the spatial dispersion in silicon can be derived from this tensor and is given by\cite{BurnettJohnH2002Sosb}:
\begin{equation}
    \label{eqSD}
    \Delta n_{SD} = 4 \sqrt{(l_1^2 l_2 ^2 + l_1^2 l_3 ^2 + l_2^2 l_3 ^2)^2 - 3 l_1^2 l_2^2 l_3^2}
\end{equation}
where $l_1, \space l_2, \text{ and } l_3$ are the direction cosines given by $l_1 = \sin \theta \cos \phi$,  $l_2 = \sin \theta \sin \phi$ and $l_3 = \cos \theta$.  
\begin{figure}
    \centering
    \includegraphics[width=1\linewidth]{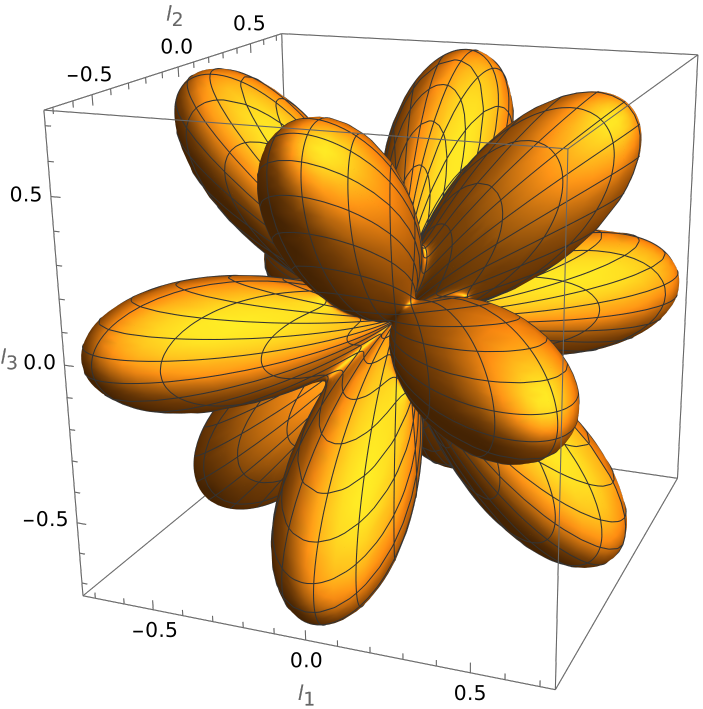}
    \caption{3D plot of the direction dependence of the birefringence introduced by considering spatial dispersion. Note the maxima at the \textlangle\num{110}\textrangle\space axes and the minima at the \textlangle\num{100}\textrangle\space and \textlangle\num{111}\textrangle\space axes. There are also saddle points present at the \textlangle\num{211}\textrangle\space axes. The birefringence has been normalised to a maximum of 1.}
    \label{fig:sdpattern}
\end{figure}
This is plotted in figure \ref{fig:sdpattern} where it can be seen that there are maxima at the face diagonals and minima at the face normals and corner diagonals. 

\subsection{Previous Measurements of Spatial Dispersion}
Pastrnak and Vedam made measurements of spatial dispersion in silicon in 1970 using two rectangular samples of silicon that had all faces polished\cite{pastrnak}. 
They used a polariser-analyser set-up with a chopped beam. 
They performed experiments at 1150\,nm and measured the birefringence along the \textlangle\num{110}\textrangle\space axis to be \num{5e-6}.  This was the only measurement of spatial dispersion induced birefringence until after the year 2000 since it was not an effect that was observable in most situations. 
An increase in interest in spatial dispersion was prompted by a larger than expected birefringence observed in crystal optics for UV experiments\cite{BurnettJohnH2002Sosb}. The observed spatial dispersion birefringence was very large due to the inverse square relation to wavelength.

In 2001, Burnett, Levine and Shirley carried out a series of \textit{ab initio} calculations of the spatial dispersion effect for several cubic crystals including silicon and compared them to existing measurements\cite{burnett2}. 
From both the work of Burnett \textit{et al.} and from Pastrnak and Vedam it is very clear that the wavelength dependence of the spatial dispersion effect is very high. 

In 2002, Chu  \textit{et al.}  measured spatial dispersion in silicon at \qty{1520}{\nm} with dislocation free silicon ingots\cite{CHU2002174}.  They found that the spatial birefringence was at a level of \num{3.2e-6}.  This is close to the measurements of Pastrnak and Vedam once scaled by the $1/\lambda^2$ relationship. From these measurements we can extrapolate the expected value of spatial dispersion at \qty{2000}{\nm} which is between \numrange[range-phrase =\,--\,, range-exponents= combine]{1.65e-6}{1.85e-6} in the case of the \textlangle\num{110}\textrangle\space axis. The purpose of our research was to measure the birefringence directly at \qty{2000}{\nm}, as well as to investigate using float zone silicon and non-normal incidence light.

\section{Spatial Dispersion in a Beamsplitter}
In the case of an input test mass the beam travels only along a single direction and so the test mass can be manufactured so that the optical path aligns with the \textlangle\num{100}\textrangle\space or \textlangle\num{111}\textrangle\space axis and experiences zero birefringence due to spatial dispersion.  In the case of a beamsplitter it is not possible to have both transmitted beams experience no birefringence.
\begin{figure}
    \centering
    \includegraphics[width=1\linewidth]{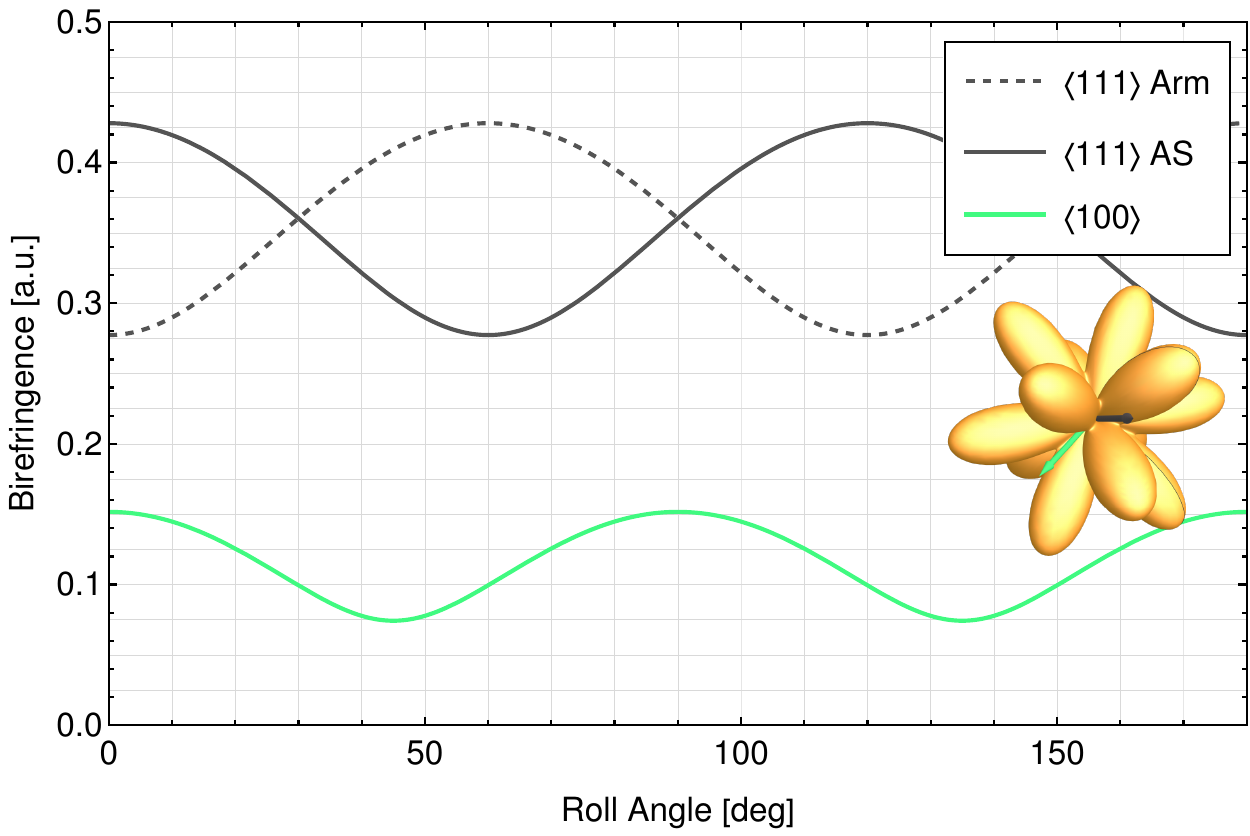}
    \caption{Change in Birefringence as a \textlangle\num{100}\textrangle\space and \textlangle\num{111}\textrangle\space silicon beamsplitter is rolled around its central axis. The birefringence has been normalised so that birefringence along the \textlangle\num{110}\textrangle\space axis is 1. The arm-transmitted beam and the antisymmetric (AS) transmitted beam in the case of  \textlangle\num{100}\textrangle\space experience the same birefringence due to the 4-fold symmetry. Note that the birefringence is lowest at roll angle of \ang{45}. The two beams in a\textlangle\num{111}\textrangle\space beamsplitter experience different birefringence levels due to the 3-fold symmetry. Inset shows the \textlangle\num{100}\textrangle\space axis as a green arrow and the \textlangle\num{111}\textrangle\space axis as a grey arrow}
    \label{fig:sd100}
\end{figure}
The incoming transmitted beam through a \ang{45} silicon beamsplitter will be refracted to \ang{12} (\qty{0.2}{\unit{rad}}) inside the optic from the surface normal. The outgoing recombined beam will be transmitted through the substrate at \ang{12} to the other side of normal, making a \ang{24} (\qty{0.4}{\unit{rad}}) difference between incoming and outgoing beams. 
This means that a \textlangle\num{100}\textrangle\space oriented silicon optic will show a spatial dispersion effect of between \qtyrange[range-phrase= --, range-units= single]{7}{15}{\%} of the maximum spatial dispersion induced birefringence, see figure \ref{fig:sd100}. 
A \textlangle\num{111}\textrangle\space optic will have more severe birefringence with spatial dispersion-induced birefringence between \qtyrange[range-phrase= --, range-units= single]{28}{43}{\%}of the maximum level. This is further complicated by the threefold symmetry of the \textlangle\num{111}\textrangle\space axis which means that when the ingoing beam experiences a birefringence of \qty{28}{\%} the outgoing beam will experience a birefringence of \qty{43}{\%} of the maximum.  

\begin{figure}
    \centering
    \includegraphics[width=1\linewidth]{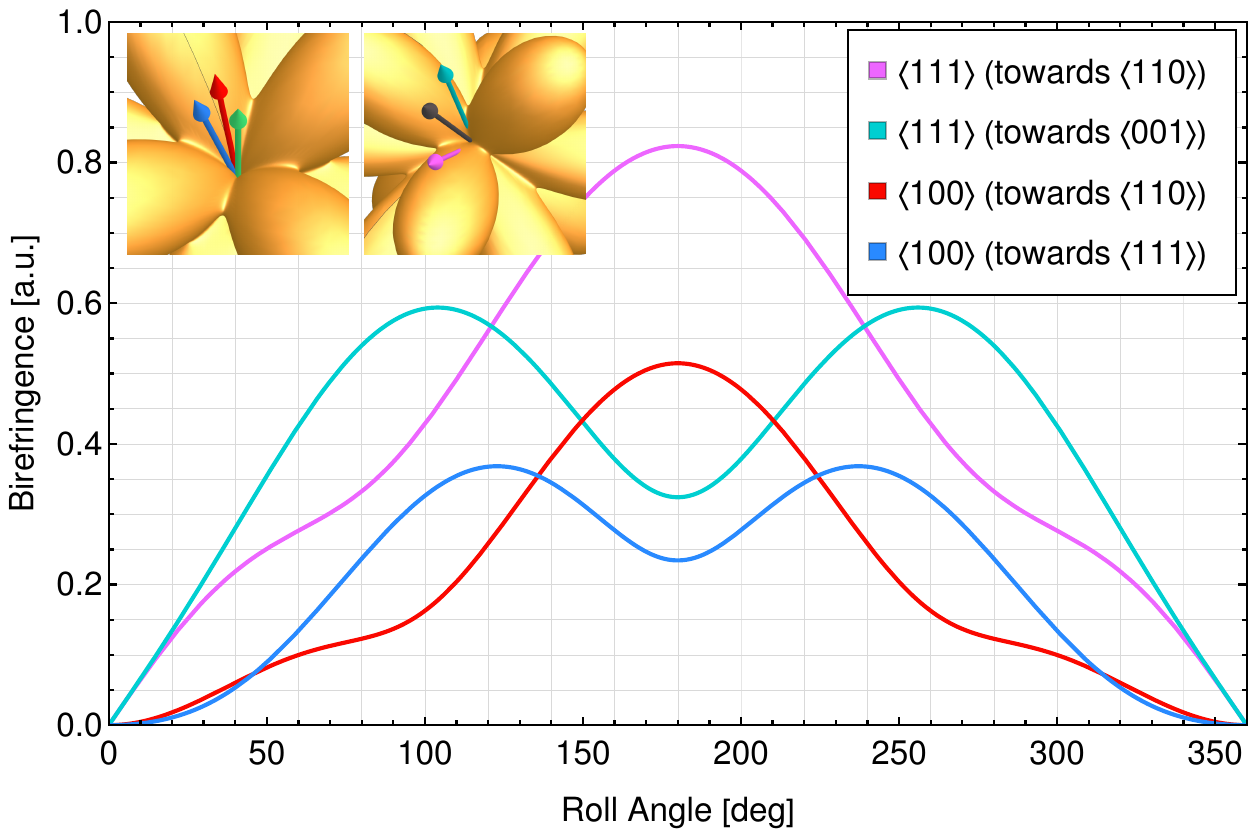}
    \caption{Birefringence as a function of roll angle in a beamsplitter that has been cut so that the one of the transmitted beams will experience no spatial dispersion contribution to the birefringence. The cuts have been made at \ang{12} from the specified crystal axis. There are two  cases for each zero-birefringence axis: for a \textlangle\num{111}\textrangle\space beamsplitter  the cut can be made towards the \textlangle001\textrangle\space axis or towards the \textlangle\num{110}\textrangle\space axis and for a \textlangle\num{100}\textrangle\space beamsplitter  the cut can be made towards the \textlangle\num{111}\textrangle\space axis or towards the \textlangle\num{110}\textrangle\space axis. The incoming beam and outgoing beam will be separated by a roll angle of \ang{180} in this figure. The axis that the cuts have been made on has been shown in the insets with the arrows representing the optics central axis. The colour of line on the plot corresponds to the colour of the arrow with the green and grey arrows representing the \textlangle\num{100}\textrangle\space and \textlangle\num{111}\textrangle\space axes respectively}
    \label{fig:cut}
\end{figure}

It would be possible to produce an optic that has been cut so that one  of the beams is along one of the zero birefringence axes. 
However, this comes at a trade-off since the other transmitted beam will experience much higher birefringence. For such a design it would be necessary to consider if a larger birefringence for the incoming beam  (in power recycling cavity) is acceptable to allow no spatial dispersion-induced birefringence for the outgoing beam (in signal recycling cavity), or vice versa. An exploration of some possible cuts is presented in figure \ref{fig:cut}.  The incoming beam will traverse the beamsplitter at a roll angle of \ang{0} in figure \ref{fig:cut}, while the outgoing beam with traverse the beamsplitter at a roll angle of \ang{180}. The \textlangle\num{100}\textrangle\space cut towards the \textlangle\num{111}\textrangle\space axis has the minimum birefringence for the outgoing beam with \qty{23}{\%} of the maximum birefringence.

\section{Methodology}
\subsection{Sample Set-up and Control}
\begin{figure}
    \centering
    \includegraphics[width=0.6\linewidth]{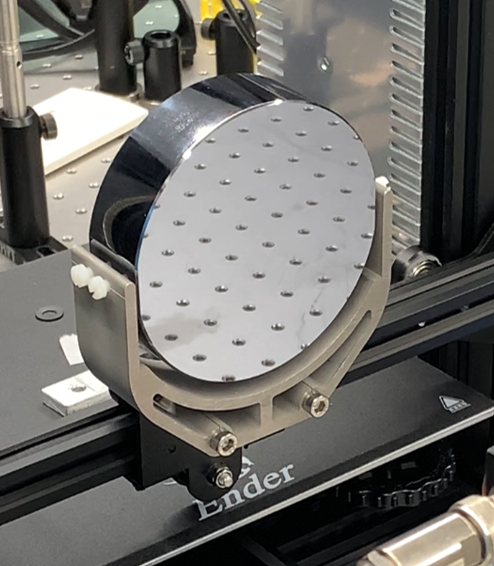}
    \caption{\textlangle\num{100}\textrangle\space float zone silicon sample used in these measurements.}
    \label{fig:sample}
\end{figure}

The sample of silicon used for these measurements was a cylindrical piece of float-zone silicon \qty{10}{\cm} in diameter and \qty{3}{\cm} thick (see figure \ref{fig:sample}). 
The central axis was collinear with the \textlangle\num{100}\textrangle\space crystallographic axis. 
The front and back surfaces are uncoated and have been polished with the front surface having a radius of curvature of \qty{44}{m} and the back surface being planar with a wedge of \ang{3}. The sample has flat sides with a pair of holes in each side which will be used to anchor the suspension points. The sample was previously used in measurements of stress induced birefringence to evaluate the suitability of float-zone silicon as an input test mass substrate\cite{vahid}. 

\begin{figure}
    \centering
    \includegraphics[width=\linewidth]{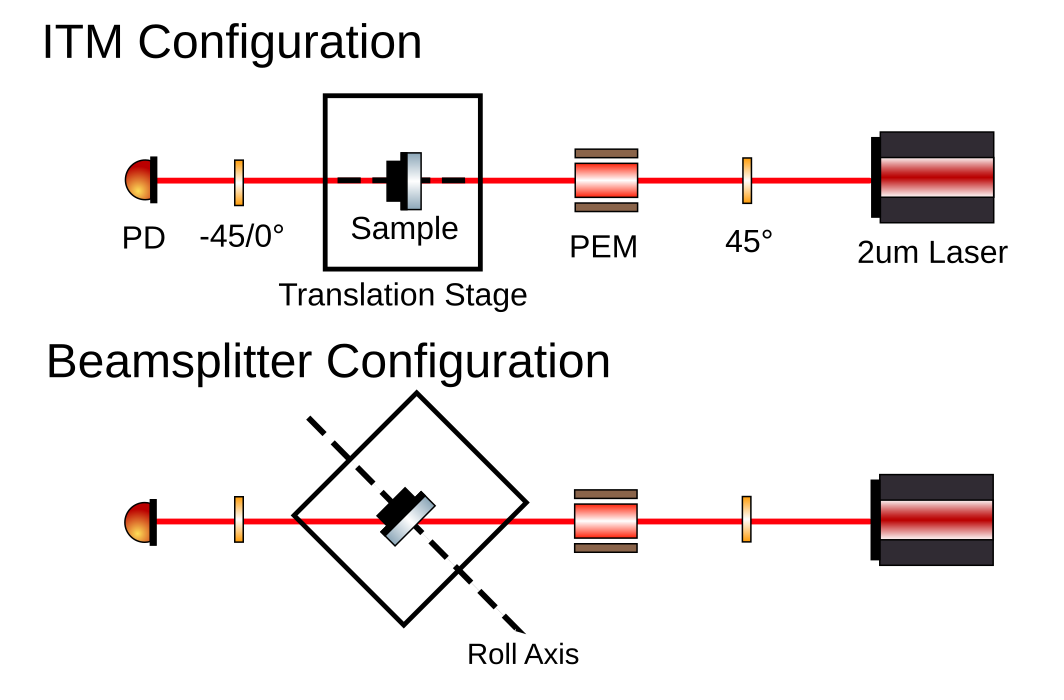}
    \caption{Experimental set up showing the difference between the measuring the birefringence in a beamsplitter and the background birefringence caused by internal and external stresses. }
    \label{fig:set_up}
\end{figure}

\begin{figure}
    \centering
    \includegraphics[width=0.6\linewidth]{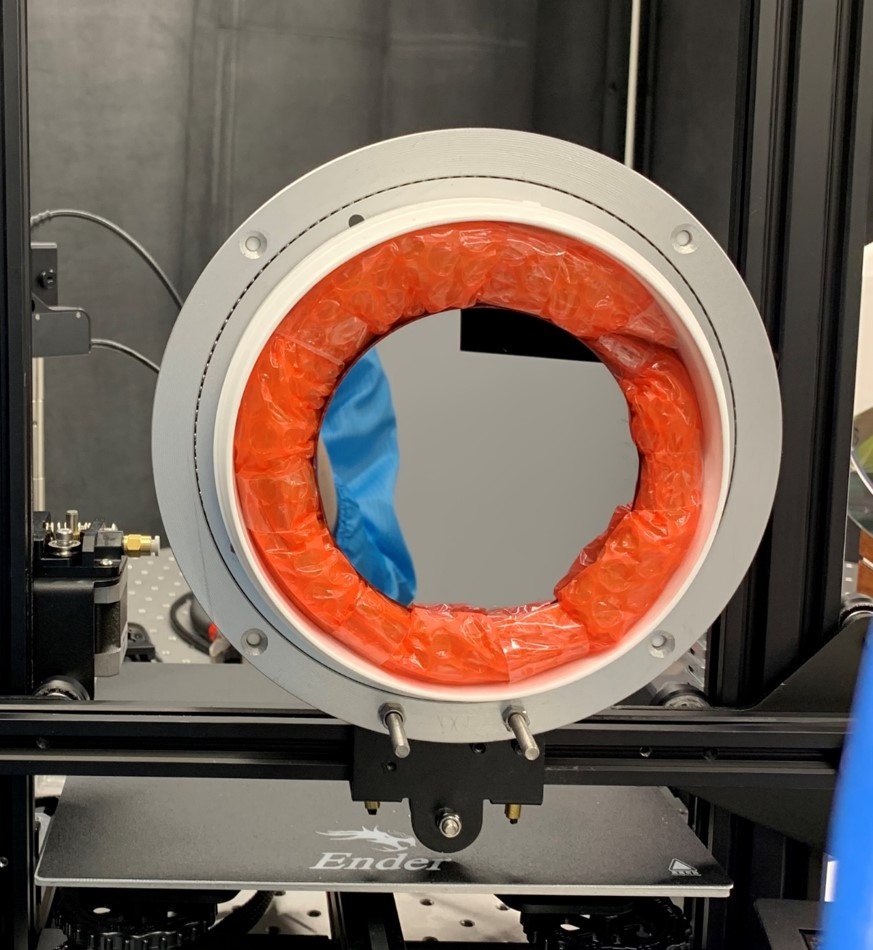}
    \caption{Silicon sample placed in rotation mount and mounted onto the translation stage. }
    \label{fig:mount}
\end{figure}

In order to investigate the birefringence in the beamsplitter case the test mass  was rotated so that the central axis of the sample was at a \ang{45} angle to the propagating beam (figure \ref{fig:set_up}). To reduce the effect of stress the sample was placed in a soft mount that was designed to produce an even stress distribution and to ensure that the stress distribution would not change between repeat measurements. The mount, shown in figure \ref{fig:mount}, was designed to rotate the test mass around the central axis, this was necessary to characterise the spatial dispersion induced birefringence. This mount consists of an inner and outer ring made of aluminium which can rotate with respect to one another. The mount contains a layer of foam wrapped in plastic to hold the sample.  The rotation of the mount was adjusted by hand between measurements. The mount was then attached to the translation stage built from a modified 3D printer. The translation stage was oriented such that the plane of motion intersected the beam at a \ang{45} angle. Control of the translation stage was via a python script to send commands to a RaspberryPi running an OctoPi server which was connected to the translation stage.
\subsection{Optical Set-up}
The optical setup for measuring birefringence is based on that described by Wang and Oakberg\cite{WangBaoliang1999Anif} which uses two polarisers and a photolelastic modulator (PEM). The beam passes through the first polariser which is set to \ang{45} to the slow axis of the PEM. It then travels through the PEM which modulates the ellipticity of the light at \qty{50}{\kHz}. The light then travels through the sample and the birefringence inherent in the sample modifies ellipticity. The second polariser is then set to 0\,degrees and the beam is measured using a photodetector producing channel 1 (ch1) data. The measurement is repeated with the second polariser set to \ang{45}, producing channel 2 (ch2) data. The second polariser is mounted on a Thorlabs motorised rotation mount (PRM1Z8) connected to a Thorlabs motor controller (KDC101). The Motor Controller can be controlled via the Thorlabs Kinesis dotNET API from a Python script. The strength of the \qty{50}{\kHz} signal (1f) is extracted using a fast Fourier transform which is then normalised by the DC signal which gives:
\begin{equation}
    \frac{V_{\text{ch1}}(1f)}{DC_{\text{ch1}}}=\delta \cos(2\rho)\frac{\sqrt{2}J_1(\Delta_0)}{1-J_0(\Delta_0)}
\end{equation}
\begin{equation}
    \frac{V_{\text{ch2}}(1f)}{DC_{\text{ch2}}}=\delta \sin(2\rho)\sqrt{2}J_1(\Delta_0)
\end{equation}
where $V_{\text{chx}}(1f)$ is the \qty{50}{\kHz} signal, $DC_{\text{chx}}$ is the DC signal, $\delta$ is the phase shift, $\rho$ is the slow axis angle and $\Delta_0$ is the modulation depth of the retardation of the PEM.  We correct for the modulation depth as we are limited to $\Delta_0 = $ \qty{1}{\unit{rad}} and so the Bessel function $J_1(\Delta_0) \neq 1$ .  
We can then define  $R_{\text{ch1}}$ and $R_{\text{ch2}}$ as:
\begin{equation}
    R_{\text{ch1}}=\frac{V_{\text{ch1}}(1f)}{DC_{\text{ch1}}}\frac{1-J_0(\Delta_0)}{2J_1(\Delta_0)}\sqrt{2} = \delta \cos(2\rho)
\end{equation}
\begin{equation}
    R_{\text{ch2}} = \frac{V_{\text{ch2}}(1f)}{DC_{\text{ch2}}}\frac{1}{2J_1(\Delta_0)} \sqrt{2} = \delta\sin(2\rho)
\end{equation}

which we can use to find the phase shift and angle:
\begin{equation}
    \delta = \sqrt{(R_{\text{ch1}})^2 +(R_{\text{ch2}})^2}
\end{equation}
\begin{equation}
    \rho = \frac{1}{2}\arctan(\frac{R_{\text{ch2}}}{R_{\text{ch1}}})
\end{equation}
From which we calculate the birefringence as a difference in refractive index between the fast and slow axis as:
\begin{equation}
    \Delta n = \frac{\lambda}{2\pi d}\delta
\end{equation}
where $\lambda$ is the laser wavelength and $d$ is the optic thickness.

\subsection{Data Processing}
The measurements are taken using a National Instruments USB X series data acquisition box using the National Instruments Python API. The \qty{50}{\kHz} signal is extracted using a Welch spectral density estimate, calculations in Eq. 11-13 are computer and the results are plotted. These functions are all controlled from a single python script running on the Lab PC which means that the process is almost entirely automated. For the \num{10000} point scans included in this paper the runtime was around \qty{24}{\unit{hrs}}. Each point measurement consisted of a \qty{3}{\s} measurement at \qty{1.4}{\unit{Msample}\,s^{-1}}. This was experimentally found to be a good sample rate and time combination for recovering known signals.

\section{Results}

\begin{figure*}[!]
    \centering
    \includegraphics[width=0.95\textwidth]{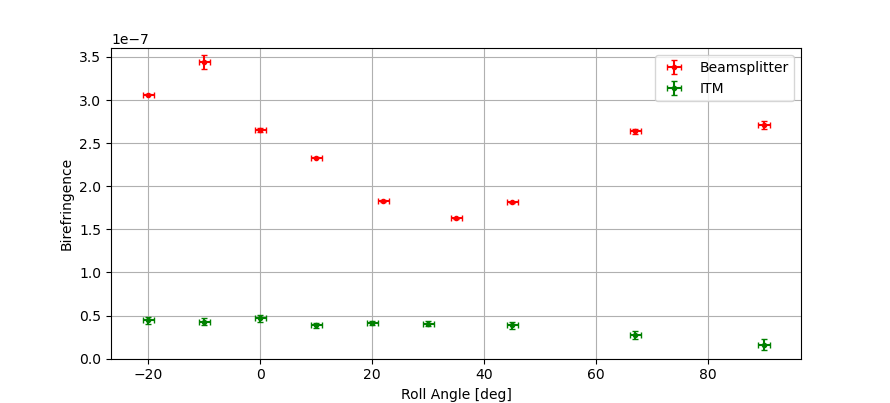}
    \caption{Measured averages of test mass central region. Points in red show the test mass in a beamsplitter configuration and the green shows measurements of the test mass in an ITM configuration. The ITM configuration gives a background birefringence due to external and internal stress of the test mass. It can be seen that this background birefringence does not have a strong angular dependence. The beamsplitter configuration demonstrates a clear minimum at \ang{35} and maximum at \ang{-10}.  }

    \label{fig:measdata}
\end{figure*}

\begin{figure*}[!]
    \centering
    \includegraphics[width=0.95\textwidth]{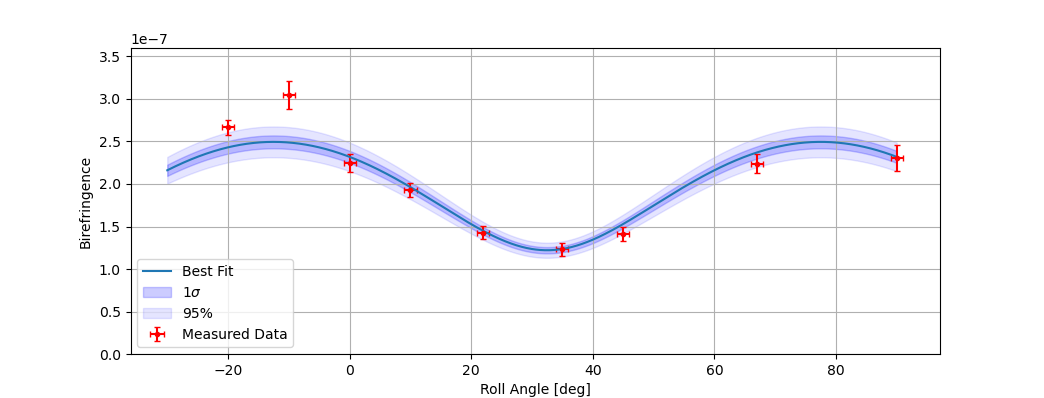}
    \caption{Measured birefringence with background birefringence subtracted. The residula represents the spatial dispersion induced birefringence.  A one parameter fit (in blue) to the data is used to estimate the maximum spatial dispersion induced birefringence.  The shaded bands give the standard error and 95\% confidence interval of the fit value for the amplitude of the maximum birefringence. The fit was carried out on the data between \ang{-20} and \ang{90}. This gave a value of \num{1.64(0.05)e-6} for the maximum birefringence.}
    \label{fig:fit_data}
\end{figure*}

We conducted nine measures at different roll angles from \ang{-20} to \ang{90}. 
Angles are referenced from the mounting orientation based on the suspension points. These scans were completed over a 100x100 point grid, which allowed for very high spatial resolution (see figure \ref{fig:scan_example} for an example of a scan, all scans are presented in the appendix. 
We also conducted 9 measurements of the background birefringence of the test mass from \ang{-20} to \ang{90} with the sample surface normal to the incident light which allowed us to characterise the stress introduced by the new mount. In order to compare measurements an average was taken as a box of 40x40 points in the centre of the test mass (figure \ref{fig:measdata}). 
The change in birefringence with the roll angle shows a sinusoidal pattern which is consistent with theory. The minimum occurs at 35\degree at a value of \num{1.63(0.05)e-7} and at a maximum of \num{3.44(0.12)e-7} at \ang{-10}.  This produces an angular separation of minimum and maximum of  \ang{45}, as expected. 
The background birefringence level was measured to be between \numrange[range-phrase =\,--\,, range-exponents= combine]{1.6e-8}{4.7e-8} and was subtracted from the measurements.  The residual is fitted to equation \ref{eqSD} which gives us a value for the maximum spatial dispersion induced birefringence (birefringence in \textlangle\num{110}\textrangle\space direction) of \num{1.64(0.05)e-6} (see figure \ref{fig:fit_data}).  

This is consistent with the measurements made by Pastrnak and Vedam using the $\frac{1}{\lambda^2}$ relation\cite{pastrnak} which predict a maximum birefringence at \qty{2000}{\nm} of \numrange[range-phrase =\,--\,, range-exponents= combine]{1.65e-6}{1.85e-6}.  The measurements made by Chu \textit{et al.} are  difficult to compare as no experimental error is given. The fitting also suggests that the \textlangle010\textrangle\space and \textlangle001\textrangle\space crystal axes are rolled \qty{13(2)}{\degree} clockwise with respect to the geometric axes as viewed facing the front surface. This is likely due to imprecision in the machining process and is of no great concern for this optics intended purpose as an input test mass. However, if this optic were intended as a silicon beamsplitter this may prove a severe error, greatly increasing birefringence because of the importance for close alignment of the crystal axes with the polarisation. These measurements demonstrate that silicon could be used as a beamsplitter if the input polarisation is within 4 degrees of the optic axes. This demonstrates the importance of well defined and consistent crystal axes throughout the optic. Visible in all of the measurements are the high birefringence artefacts that occur in the upper left quadrant of the test mass noted in\cite{vahid}.  These can be seen to rotate with the test mass rotation which demonstrates that they are of physical origin.  
 \begin{figure}
    \includegraphics[width=\linewidth]{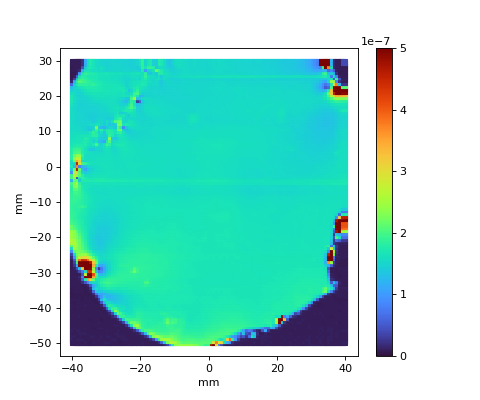}
    \caption{Measurement of spatial dispersion birefringence with the lowest average birefringence. The sample is at an angle of \ang{45} with the incident light. The roll angle of the sample is \ang{35}.  The features in the top left quadrant are real and observed to rotate with the the sample (see appendix). The horizontal striations are not real and are an artefact of the measurement}
    \label{fig:scan_example}
    \end{figure}


\section{Conclusion}
We demonstrated that float-zone silicon is a viable beamsplitter substrate for future gravitational wave detectors.
Care must be taken to ensure alignment of the crystal axis of the optic with the light polarisation. 
In this demonstration the test mass crystal axes were not aligned to the test mass geometric axes which would increase spatial dispersion induced birefringence if the optic were used as a beam splitter as manufactured. 
The spatial dispersion induced birefringence level of a \textlangle\num{100}\textrangle\space beamsplitter is \numrange[range-phrase =\,--\,, range-exponents= combine]{\approx1e-7}{3e-7}. 
For reference \num{1e-7} is an acceptable level, particularly if light polarisation is aligned to the optical axes. 
We also considered the option of cutting the optic so that the light can be transmitted along the non-birefringent axes. For a \ang{45} optic the beam will propagate at \ang{12} away from the normal, this means that the light transmitted towards the arm and the light transmitted towards the antisymmetric port will separated by \ang{24}. So simply changing the cut direction can not eliminate spatial dispersion induced birefringence so one beam must always experience birefringence.  It may be possible to introduce compensation optics to remove the effect of birefringence.  The spatial dispersion effect is a homogeneous effect so it is simpler to compensate compared to a complex birefringence pattern.

This study also reinforces the importance of high-resolution scans of any silicon optics. A scan of low resolution or only a single point would not have measured the high-birefringence points that can be seen close to the edge of the test mass. These were likely introduced during the manufacturing of the crystal. Our measurements also offer support for using \textlangle\num{100}\textrangle\space silicon and \qty{2000}{\nm} light in the detector. This is because a \textlangle\num{111}\textrangle\space beamsplitter will produce a larger birefringence due to the \textlangle\num{111}\textrangle\space zero birefringence point being a zero crossing while the \textlangle\num{100}\textrangle\space is a local minimum.  The inverse square dependence on wavelength supports \qty{2000}{\nm} over \qty{1550}{\nm} which will experience a \qty{60}{\%} increase in birefringence due to spatial dispersion.  Ultimately we have demonstrated that silicon is a plausible beamsplitter substrate material. However, additional manufacturing constraints are required to limit spatial dispersion induced birefringence. 
\section{Acknowledgments}
The authors would like to thank the members of the UWA gravitational wave instrumentation group for their useful and insightful feedback over the course of this project. The authors would also like to thank Professor Li Ju and Felix Wojcik for their help in constructing the rotation mount. This work was supported by the Australian Research Council Centre of Excellence for Gravitational Wave Discovery (CE230100016). 

\bibliography{references}

\clearpage
\appendix
\label{app:appendixa}

\section{Birefringence Scans}

\begin{figure}[h!]
    \begin{subfigure}{\textwidth}
    \includegraphics[width=0.7\textwidth]{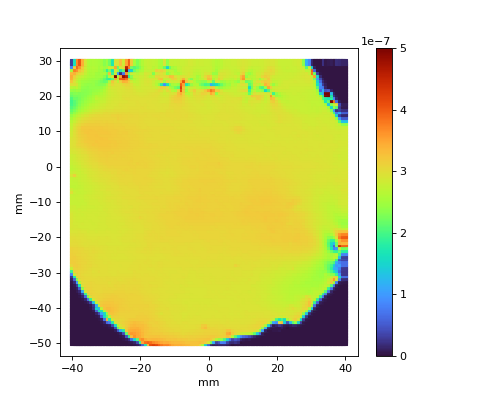}
    \caption{$-20\degree$ Roll}
    \label{fig:neg20}
    \end{subfigure}
        \begin{subfigure}{\textwidth}
        
   \includegraphics[width=0.7\textwidth]{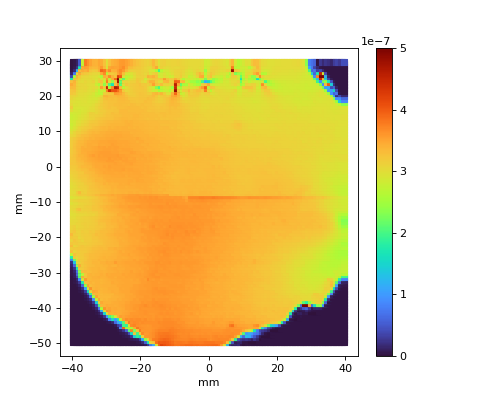}
    \caption{$-10\degree$ Roll}
    \label{fig:neg10}
    \end{subfigure}
   \end{figure}
       \begin{figure}\ContinuedFloat
   \centering
        \begin{subfigure}{\textwidth}
    \includegraphics[width=0.7\textwidth]{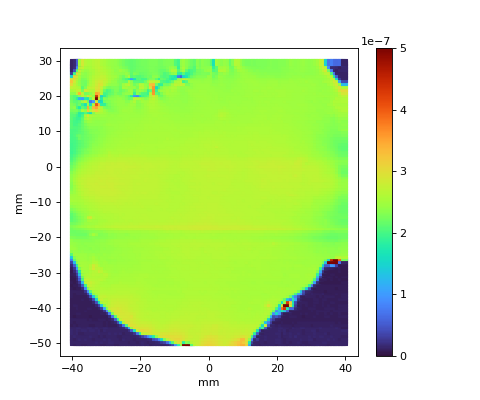}
    \caption{$0\degree$ Roll}
    \label{fig:0}
    \end{subfigure}
        \begin{subfigure}{\textwidth}
    \includegraphics[width=0.7\textwidth]{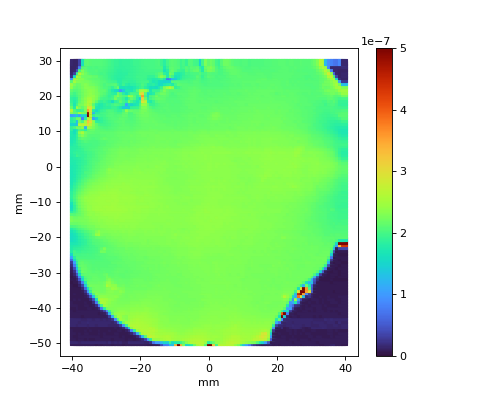}
    \caption{$10\degree$ Roll}
    \label{fig:10}
    \end{subfigure}
    \end{figure}
    \begin{figure}\ContinuedFloat
   \centering
            \begin{subfigure}{\textwidth}
    \includegraphics[width=0.7\textwidth]{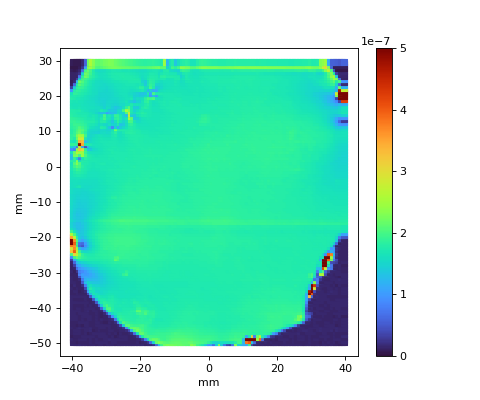}
    \caption{$22\degree$ Roll}
    \label{fig:22}
    \end{subfigure}
    
            \begin{subfigure}{\textwidth}
    \includegraphics[width=0.7\textwidth]{35_map_bs.png}
    \caption{$35\degree$ Roll}
    \label{fig:35}
    \end{subfigure}
    \end{figure}
    \begin{figure}\ContinuedFloat        
    \centering
             
    \begin{subfigure}{\textwidth}
    \includegraphics[width=0.7\textwidth]{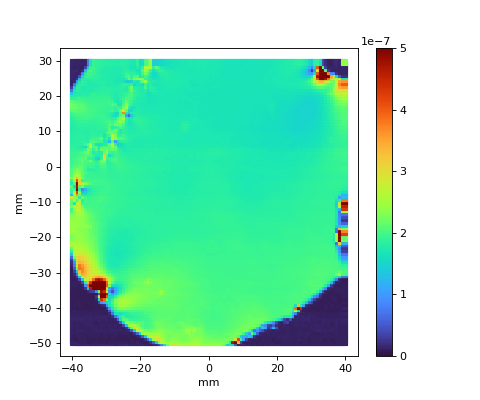}
    \caption{$45\degree$ Roll}
    \label{fig:45}
    \end{subfigure}

    \begin{subfigure}{\textwidth}
    \includegraphics[width=0.7\textwidth]{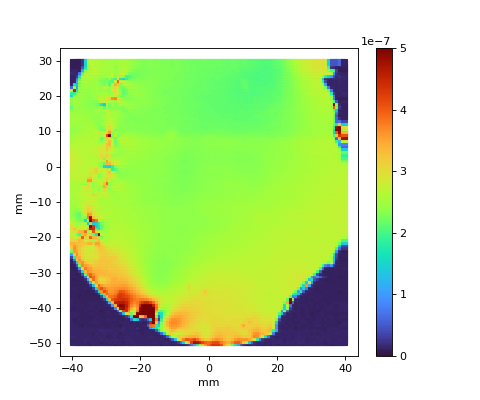}
    \caption{$67\degree$ Roll}
    \label{fig:67}
    \end{subfigure}
    \end{figure}
    \begin{figure}\ContinuedFloat        
\captionsetup{width=\textwidth}
                     \begin{subfigure}{\textwidth}
    \includegraphics[width=0.7\textwidth]{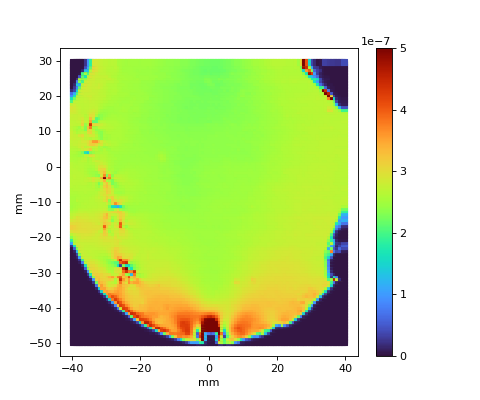}
    \caption{$90\degree$ Roll}
    \label{fig:neg90}
    \end{subfigure}
  
    \caption{Ten thousand point scans of the birefringence in the silicon test mass in a beamsplitter orientation as it is rolled around its central axis from $-20\degree$  to $90\degree$ with respect to its mounting orientation. The horizontal artefacts present are not real an only an effect of the measurement apparatus. The irregular shape around the sides of the test mass is due to the beam being blocked by the foam that surrounds it. The high birefringence that occurs directly over the machined holes should not be taken as real. }
    \label{fig:all}
\end{figure}

\end{document}